\documentclass[aps,prl,preprintnumbers,showpacs,nofootinbib]{revtex4}
\usepackage{graphicx}% Include figure files
\usepackage{dcolumn}% Align table columns on decimal point
\usepackage{bm}% bold math
\begin{document}
\newcommand{\newc}{\newcommand}
\newc{\ra}{\rightarrow}
\newc{\lra}{\leftrightarrow}
\newc{\beq}{\begin{equation}}
\newc{\eeq}{\end{equation}}
\newc{\barr}{\begin{eqnarray}}
\newc{\earr}{\end{eqnarray}}
%%%%%%%%%%%%%%%%%%%%%%%%%%%%%%%%%%%%%%%%%%%
\newcommand{\Od}{{\cal O}}
\newcommand{\lsim}   {\mathrel{\mathop{\kern 0pt \rlap
  {\raise.2ex\hbox{$<$}}}
  \lower.9ex\hbox{\kern-.190em $\sim$}}}
\newcommand{\gsim}   {\mathrel{\mathop{\kern 0pt \rlap
  {\raise.2ex\hbox{$>$}}}
  \lower.9ex\hbox{\kern-.190em $\sim$}}}
  \def\rpm{R_p \hspace{-0.8em}/\;\:}
%\preprint{APS/123-QED}
%\date{\today}
\title {SUPERNOVA DETECTION VIA A NETWORK OF NEUTRAL CURRENT SPHERICAL TPC'S}
%\title {EXPLOITING  TPC DETECTORS AND THE COHERENT  NEUTRAL CURRENT INTERACTION
%FOR DETECTING  SUPERNOVA NEUTRINOS}
%
%
%\toctitle{ Neutrinos \protect\newline in a Spherical Box}
% allows explicit linebreak for the table of content
%
%
%\titlerunning{NEUTRINOS IN A SPHERICAL BOX}
% allows abbreviation of title, if the full title is too long
% to fit in the running head
%
\author{ J.D. Vergados$^{1,2}$ and Y. Giomataris$^{3}$}
\affiliation{$^1$University of Ioannina, Ioannina, GR 45110, Greece\\
$^2$RCNP, Osaka University, Ibaraki,Japan.\\
 $^3$ CEA, Saclay, DAPNIA, Gif-sur-Yvette, Cedex,France.
%\\E-mail:Vergados@cc.uoi.gr
}
%\maketitle              % typesets the title of the contribution
%\begin{frontmatter}
\vspace{0.5cm}
\begin{abstract}
The coherent contribution of all neutrons in neutrino nucleus scattering due to the neutral current
offers a realistic prospect of detecting supernova neutrinos. For a
typical supernova at $10$ kpc, about 1000 events are expected using
 a spherical gaseous detector  of radius 4 m and
employing Xe gas at a pressure of 10 Atm. We propose a world wide network of several such simple,
stable and low cost supernova detectors with a running time of a few centuries.
\end{abstract}
%\end{document}
%\begin{keyword}
%%%%%%%%%%%%%%%%%%%%%%%%%%%%%%%%%%%%%%%%%%%%%%%%%%%%%%%%%%%%%%%%%%%%%
%\date{\today}
\maketitle
%\end{keyword}
%\end{frontmatter}
%%%%%%%%%%%%%%%%%%%%%%%%%%%%%%%%%%%%%%%%%%%%%%%%%%%%%%%%%%%%%%%%%%%%%
%%%%%%%%%%%%%%%%%%%%%%%%%%%%%%%%%%%%%%%%%%%%%%%%%%%%%%%%%%%%%%%%%%%%%
\section{Introduction.}
In a typical supernova an energy of about $10^{53}$ ergs is released in the form of neutrinos
\cite{BEACFARVOG},\cite{SUPERNOVA}. These neutrinos
are emitted within an interval of about $10$ s after the explosion and they travel to Earth undistorted, except that,
on their way to Earth, they may oscillate into other flavors. The phenomenon of neutrino oscillations is by
 now established by the observation of
 atmospheric neutrino oscillations \cite{SUPERKAMIOKANDE} interpreted as
 $\nu_{\mu} \rightarrow \nu_{\tau}$ oscillations, as well as
 $\nu_e$ disappearance in solar neutrinos \cite{SOLAROSC}. These
 results have been recently confirmed by the KamLAND experiment \cite{KAMLAND},
 which exhibits evidence for reactor antineutrino disappearance. Thus for traditional detectors
 relying on the charged current
 interactions the precise event rate may depend critically on the specific properties of the neutrinos. The
time integrated spectra in the case of charged current detectors, like the SNO experiment,
depend on the neutrino oscillations \cite{TKBT03}.
 An additional problem is the fact that the charged current cross sections depend on the details of the
 structure of the nuclei involved.
With neutral current detectors one exploits the fact that the vector component of the current can lead
 to coherence, i.e. an additive contribution of all neutrons in the nucleus (the proton component is tiny). Furthermore the deduced neutrino fluxes do not depend on the  neutrino oscillation parameters (e.g. the mixing angles). Even in our case, however, the obtained rates depend on the assumed characteristic temperature for each flavor, see sec. \ref{sec.supernova}.

% During the last years various detectors aiming at detecting recoiling nuclei have been developed
%in connection with  dark matter
%searches \cite{CDMALL} with thresholds in the $10$ keV region.
Recently it has become feasible to detect neutrinos by measuring
 the recoiling nucleus  and employing gaseous detectors with much lower threshold energies. Thus
 one is able to explore the advantages offered by the neutral current interaction, exploring ideas put forward more than a decade  ago \cite{DKLEIN}.
A description of the NOSTOS project and details of the spherical TPC detector are given in \cite{NOSTOS1}.
 We have built a spherical prototype 1.3 m in diameter
which is described in \cite{NOSTOS2}. The outer vessel is made of pure Cu (6 mm thick)
allowing to sustain pressures up to 5 bar. The inner detector is just a small sphere, 10 mm
in diameter \cite{GRRC} and  developments are currently under way
to build a spherical TPC detector using new
technologies. First tests were performed by filling
the volume with argon mixtures and are quite promising. High gains are easily obtained and
 the signal to noise is large enough for sub-keV threshold. The whole system looks stable
 and robustaaaaaaaaaaaaa made of stainless
steel as a proportional counter located at the center of
curvature of the TPC.
% there are no problems associated with uncertainties in the flux of any neutrino flavor due to oscillations.
Furthermore this interaction, through its vector component, can lead
 to coherence, i.e. an additive contribution of all nucleons in the nucleus. Since the vector  contribution of the
 protons is tiny, the coherence is mainly due to the neutrons of the nucleus.

 In this paper we will derive the amplitude for the differential neutrino nucleus coherent cross section. Then
 we will  estimate
 the expected number of events for all the noble gas targets. We will show that these results can be exploited
by a network of small and relatively cheap spherical TPC detectors placed in various parts of the world
(for a description of the apparatus see our
earlier work \cite{NOSTOS1}). The operation of such devices as a network will minimize the background
problems. There is no need to go underground,
but one may have to go sufficiently deep underwater to balance the high pressure of the gas target. Other types
of detectors have also been proposed \cite{MICROPATTERN},\cite{TWOPHASE}.\\
Large gaseous volumes are easily obtained by employing long drift technology (i.e TPC) that can
provide massive targets by increasing the gas pressure. Combined with an adequate amplifying
structure and low energy thresholds, a three-dimensional reconstruction of the recoiling particle, electron
 or nucleus, can be obtained. The use of new micropattern detectors and especially the novel Micromegas
\cite{GIOMA96} provide excellent spatial and time accuracy that is a precious tool for pattern recognition and
 background rejection \cite{CG01},\cite{GORO05}.
The virtue of using such large gaseous volumes and the new high precision microstrip gaseous detectors
has been recently discussed in a dedicated workshop \cite{WORKSHOP04} and their relevance for low energy neutrino
physics and dark matter detection has been widely recognized. Such  low-background low-energy threshold
systems are actually successfully used in the CAST \cite{AALSETH}, the solar axion experiment,
and are under development
for several low energy neutrino or dark matter projects
 \cite{NOSTOS1},\cite{DARKMATTER}.
\section{\label{sec:level2} Standard and non standard weak interaction}
The standard neutral current  left handed weak interaction can be cast in the form:
\begin{equation}
{\cal L_q}=-\frac{G_F}{\sqrt{2}}
\left[ \bar\nu_\alpha \gamma^\mu (1-\gamma^5)\nu_\alpha \right]
\left[ \bar q\gamma_\mu (g_V(q)-g_A(q)\gamma^5) q \right]
\label{weak1}
\end{equation}
(diagonal in flavor space).
% with
%\begin{eqnarray}
%g_V(u)&=&\frac{1}{2}-\frac{4}{3} \sin^2{\theta_W}~~,~~g_A(u)=\frac{1}{2}~;
\nonumber\\
%g_V(d)&=&-\frac{1}{2}+\frac{2 }{3}\sin^2{\theta_W}~~,~~g_A(u)=-\frac{1}{2}
%\label{weak2}
%\end{eqnarray}
At the nucleon level we get:
\begin{equation}
{\cal L_q}=-\frac{G_F}{\sqrt{2}}
\left[ \bar\nu_\alpha \gamma^\mu (1-\gamma^5)\nu_\alpha \right]
\left[ \bar N\gamma_\mu (g_V(N)-g_A(N)\gamma^5) N \right]
\label{weak3}
\end{equation}
 with
\begin{equation}
g_V(p)=\frac{1}{2}-2 \sin^2{\theta_W}\simeq 0.04~~,~~g_A(p)=1.27 \frac{1}{2}~;~
g_V(n)=-\frac{1}{2}~~,~~g_A(n)=-\frac{1.27}{2}
\label{weak4}
\end{equation}
%\begin{eqnarray}
%g_V(p)&=&\frac{1}{2}-2 \sin^2{\theta_W}\simeq 0.04~~,~~g_A(p)=1.27 \frac{1}{2}~;\nonumber\\
%\label{weak4}
%\end{eqnarray}
Beyond the standard level one has further interactions which need not be diagonal in flavor space. Thus
\begin{eqnarray}
& &g_V(q)-g_A(q)\gamma^5 \rightarrow
\nonumber\\
& &\left( g^{SM}_V(q)-g^{SM}_A(q)\gamma^5 \right)
\delta_{\alpha \beta}+\left( \lambda^{qL}\delta_{\alpha \beta}+\epsilon^{qL}_{\alpha \beta} \right)(1-\gamma^5)
\nonumber\\
& &\left[ \bar\nu_\alpha \gamma^\mu (1-\gamma^5)\nu_\alpha \right]\rightarrow
\left[ \bar\nu_\alpha \gamma^\mu (1-\gamma^5)\nu_\beta \right]
\label{weak5}
\end{eqnarray}
Furthermore at the nucleon level
\begin{eqnarray}
g_V(N)-g_A(N)\gamma^5 \rightarrow& &
\left( g^{SM}_V(N)-g^{SM}_A(N)\gamma^5 \right) \delta_{\alpha \beta}
\nonumber\\
& & + \left( \lambda^{NL}\delta_{\alpha \beta}+\epsilon^{NL}_{\alpha \beta} \right)(1-1.27 \gamma^5)
\nonumber\\
\left[ \bar\nu_\alpha \gamma^\mu (1-\gamma^5)\nu_\alpha \right]\rightarrow & &
\left[ \bar\nu_\alpha \gamma^\mu (1-\gamma^5)\nu_\beta \right]
\label{weak6}
\end{eqnarray}
with
\begin{eqnarray}
\lambda^{pL}&=&2 \lambda^{uL}+\lambda^{dL}~,~\lambda^{nL}=\lambda^{dL}+ 2\lambda^{dL}~,
\nonumber\\
& &\epsilon^{pL}_{\alpha \beta}=2\epsilon^{uL}_{\alpha \beta}+\epsilon^{dL}_{\alpha \beta}~,
~\epsilon^{nL}_{\alpha \beta}=\epsilon^{uL}_{\alpha \beta}+ 2\epsilon^{dL}_{\alpha \beta}
\label{weak7}
\end{eqnarray}
In the above expressions $\lambda^{qL}$ can arise, e.g.,  from radiative corrections, see e.g. PDG \cite{PDG} and $\epsilon^{qL}_{\alpha \beta}$ from R-parity violating interactions in supersymmetric models \cite{HIRSCH00}-\cite{HFVK00}.\\
Indeed since R-parity conservation has no robust
theoretical motivation one may accept an extended framework of
the MSSM with R-parity non-conservation  MSSM.
In this case the superpotential
$W$ acquires additional R-paity violating terms:
\begin{eqnarray}
  W_{\rpm} & = & \lambda_{ijk}L_{i}L_{j}E^c_{k}
  + \lambda^{\prime}_{ijk}L_{i}Q_{j}D^c_{k}\nonumber\\
 &+& \lambda^{\prime \prime}_{ijk} U^c_{i}D^c_{j}D^c_{k}+ \mu_j L_{j}H_u
    \label{rpsuperpotential}
\end{eqnarray}
Of interest to us here is the $\lambda^{\prime}_{ijk}L_{i}Q_{j}D^c_{k}$ involving first generation quarks and s-quarks,
i.e the term $\lambda^{\prime}_{\alpha 1 1} L_{\alpha}Q_{1}D^c_{1}$. From this term in four component notation we get the contribution
$$\lambda^{\prime}_{\alpha 1 1} \left (\bar{d^c_R} \nu_{\alpha L}- \bar{u^c_R} \alpha_L \right )\tilde{d}^c,~\alpha =e,\mu,~\tau$$
where $\nu_{\alpha L}=\frac{1}{2}(1-\gamma_5)\nu_{\alpha}$ etc. Thus
\begin{itemize}
\item The first term at tree level yields the interaction
\beq
 -\frac{\lambda^{\prime}_{\alpha 1 1} \lambda^{\prime}_{\beta 1 1}}{m^2_{\tilde{d}_L}}
 \bar{\nu}_{\alpha L} d^c_R  \bar{d^c_R} \nu_{\beta L}
\label{rparity1}
\eeq
By performing a Fierz transformation we can rewrite it in the form:
\beq
 \frac{1}{2}\frac{\lambda^{\prime}_{\alpha 1 1} \lambda^{\prime}_{\beta 1 1}}{m^2_{\tilde{d}_L}}
 \bar{\nu}_{\alpha L} \gamma^{\mu} \nu_{\beta L} \bar{d^c_R} \gamma_{\mu} d^c_R
\label{rparity2}
\eeq
The previous equation can be cast in the form:
\begin{eqnarray}
{\cal L}_d&=&-\frac{G_F}{\sqrt{2}} \epsilon ^d_{\alpha \beta}
\left[ \bar{\nu_{\alpha}} \gamma^\mu ( 1-\gamma^5 )\nu_\beta \right]\left[ \bar{d}\gamma_\mu ( 1-\gamma^5 )d \right ];
\nonumber\\
 \epsilon ^d_{\alpha \beta}&=& \lambda^{\prime}_{\alpha 1 1} \lambda^{\prime}_{\beta 1 1} \frac{m_W^2}{m^2_{\tilde{d}_L}}
\label{rparity3}
\end{eqnarray}
There is no such term associated with the $u$ quark, $\epsilon ^u_{\alpha \beta}=0$.
\item Proceeding in an analogous fashion the collaborative effect  of the first and second term, for $\alpha,\beta=e,\mu,\tau$, yields
the charged current contribution:
\begin{eqnarray}
{\cal L}_{du}&=&\frac{G_F}{\sqrt{2}} \epsilon ^d_{\alpha \beta}
\left[ \bar{\alpha} \gamma^\mu (1-\gamma^5)\nu_\beta \right]\left[ \bar{u}\gamma_\mu (1-\gamma^5)d \right ]
%\nonumber\\
% \epsilon ^d_{\alpha \beta}&=& \lambda^{\prime}_{\alpha 1 1} \lambda^{\prime}_{\beta 1 1} \frac{m_W^2}{m^2_{\tilde{d}_L}}
\label{rparity4}
\end{eqnarray}
\item Finally the second term, for $\alpha,\beta=e,\mu,\tau$, leads to a neutral current contribution of the charged leptons:
\begin{eqnarray}
{\cal L}_u &=&\frac{G_F}{\sqrt{2}} \epsilon ^d_{\alpha \beta}
\left[ \bar{\alpha} \gamma^\mu ( 1-\gamma^5 ) \beta \right]\left[ \bar{u}\gamma_\mu ( 1-\gamma^5 )u \right ]
%\nonumber\\
% \epsilon ^d_{\alpha \beta}&=& \lambda^{\prime}_{\alpha 1 1} \lambda^{\prime}_{\beta 1 1} \frac{m_W^2}{m^2_{\tilde{d}_L}}
\label{rparity5}
\end{eqnarray}
\end{itemize}
The above non standard flavor changing neutral current interaction have been found to play an important role in the in the infall stage
of a stellar collapse \cite{AFG05}. Furthermore precise measurements involving the neutral current neutrino-nucleus interactions may yield valuable information about the non standard interactions \cite{BMR05}. They are not, however, going to be further considered in this work.

%\section{ Elastic Neutrino nucleon Scattering}
% For low energy neutrinos the historic process neutrino-electron scattering \cite{HOOFT} \cite{REINES}
% is very useful.
%The differential cross section \cite{VogEng} takes the form
%('t Hooft and Vogel $\&$ Engel)
%\begin{equation}
%\frac{d\sigma}{dT}=\left(\frac{d\sigma}{dT}\right)_{weak}+
%\left(\frac{d\sigma}{dT}\right)_{EM} \label{elas1a}
%\end{equation}
\section {Coherent neutrino nucleus scattering}
The cross section for elastic neutrino nucleon scattering has extensively been studied
\cite{BEACFARVOG},\cite{VogEng}.

The energy of the recoiling paricle can be written in dimensionless form
as follows:
\beq
y=\frac{2\cos^2{\theta}}{(1+1/x_{\nu})^2-\cos^2{\theta}}~~,~~
y=\frac{T_{recoil}}{m_{recoil}},x_{\nu}=\frac{E_{\nu}}{m_{recoil}}
\label{recoilen}
\eeq
 The maximum  energy occurs when $\theta=0$, $y_{max}=\frac{2}{(1+1/x_{\nu})^2-1}$,
in agreement with Eq. (2.5) of ref. \cite{BEACFARVOG}.
% This relationship is plotted in  Fig. \ref{fig:yofx}.
%I what follows, whenever appropriate, we are going to average our results with the neutrino spectrum
%shown in Fig. \ref{spectrum}.
  One can invert Eq. \ref{recoilen} and get the  neutrino energy associated with a given recoil energy and
scattering angle.
From the above expressions we see that the vector current contribution, which may lead to coherence, is negligible
in the case of the protons. Thus the coherent contribution \cite{PASCHOS} may come from the neutrons and is expected to be
proportional to the square of the neutron number.
%The neutrino-nucleus scattering can be obtained from the amplitude of the neutrino nucleon scattering under
%the following assumptions:
%\begin{itemize}
%\item Employ the appropriate kinematics, i.e. those involving the elastically scattered nucleus.
%\item Ignore  effects of the nuclear form factor. Such effects, which are not expected to be very large,
% are currently under study and they will
% appear elsewhere.
%\item The effective neutrino-nucleon amplitude is obtained as above with the substitution
%$${\bf q}\Rightarrow \frac{{\bf p}}{A}~~,~~E_N \Rightarrow \sqrt{m_N^2+\frac{{\bf p}^2}{A^2}}=\frac{E_A}{A}$$
%with ${\bf q}$ the nucleon momentum and ${\bf p}$ the nuclear momentum.
%\end{itemize}
%The neutrino-nucleus cross section takes the form:
% \begin{eqnarray}
% \left(\frac{d\sigma}{dT_A}\right)_{weak}&=&\frac{G^2_F Am_N}{2 \pi}
% [(M_V+M_A)^2 \left (1+\frac{A-1}{A}\frac{T_A}{E_{\nu}} \right )
% \nonumber\\
%+ (M_V-M_A)^2
%(1&-&\frac{T_A}{E_{\nu}})^2
%\left (1-\frac{A-1}{A}\frac{T_A}{m_N}\frac{1}{E_{\nu}/T_A-1} \right )
%\nonumber\\
%&+& (M_A^2-M_V^2)\frac{Am_NT_A}{E^2_{\nu}} ]
% \label{elaswA}
%  \end{eqnarray}
 % Where $M_V$ and $M_A$ are the nuclear matrix elements associated with the vector and the axial current
 % respectively and $T_A$ is the energy of the recoiling nucleus.
% The axial current contribution vanishes for $0^+ \Rightarrow 0^+$ transitions. Anyway it is negligible
%  in front of the coherent scattering due to neutrons. Thus the previous formula is reduced to:
  The neutrino-nucleus coherent cross section takes the form:

   \begin{eqnarray}
 \left(\frac{d\sigma}{dT_A}\right)_{weak}&=&\frac{G^2_F Am_N}{2 \pi}~(N^2/4)F_{coh}(A,T_A,E_{\nu}),
\nonumber\\
& &F_{coh}(A,T_A,E_{\nu})=
  \left (1+\frac{A-1}{A}\frac{T_A}{E_{\nu}} \right )
+(1-\frac{T_A}{E_{\nu}})^2
\nonumber\\
& &\left (1-\frac{A-1}{A}\frac{T_A}{m_N}\frac{1}{E_{\nu}/T_A-1} \right )
-\frac{Am_NT_A}{E^2_{\nu}}
 \label{elaswAV}
  \end{eqnarray}

%  We see two reasons for enhancement of the cross section:
%\begin{itemize}
%\item The overall A factor due to the kinematics, which is  counteracted by the smaller
%nuclear recoil energy when compared to the nucleon recoil energy for the same neutrino energy.
%This factor will be absorbed into the energy integrals, see the function $F_{fold}(A,T,(T_A)_{th})$ below.
%\item The $N^2$ enhancement due to coherence.
%\end{itemize}
\section{Supernova Neutrinos}
\label{sec.supernova}
\begin{figure}[!ht]
 \begin{center}
 \rotatebox{90}{\hspace{1.0cm} {$\frac{dN}{d E_{\nu}}\rightarrow \frac{10^{58}}{MeV}$}}
\includegraphics[scale=0.8]{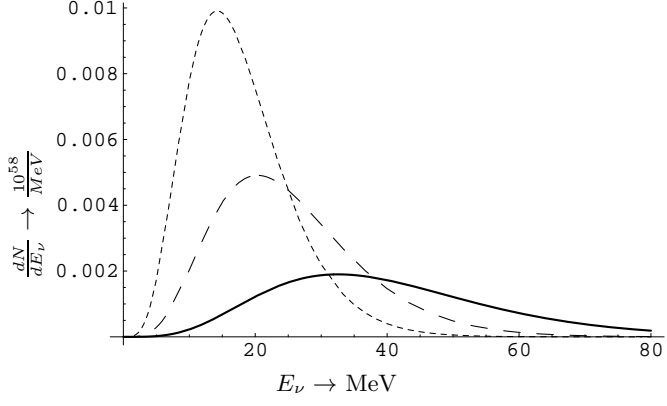}
\hspace{8.0cm}$E_{\nu} \rightarrow$ MeV
 \caption{The supernova neutrino spectrum. The short dash, long dash and continuous curve correspond
to $\nu_e,\tilde{\nu}_e$ and all other flavors respectively}
 \label{supernovasp}
 \end{center}
  \end{figure}

The number of neutrino events for a given detector depends on the neutrino spectrum and the distance of the
source. We will consider a typical case of a source which is about $10$ kpc, l.e. $D=3.1 \times 10^{22}$ cm ( of the order of the radius of the galaxy) with
an energy output of $3 \times 10^{53}$ ergs with a duration of about $10$ s.  Furthermore we will assume for simplicity that each neutrino flavor is characterized by  a
 Fermi-Dirac like distribution times its characteristic cross section and we will not consider here the more
realistic distributions, which have recently become available \cite{NUSPECTRA}.
This is adequate for our purposes. Thus:
\beq
\frac{dN}{dE_{\nu}}=\sigma(E_{\nu})\frac{E^2_{\nu}}{1+exp(E_{\nu}/T)}=\frac{\Lambda}{JT}\frac{x^4}{1+e^x}~~,
~~x=\frac{E_{\nu}}{T}
\label{nudistr}
\eeq
with
$J=\frac{31\pi^6}{252}$, $\Lambda$  a constant and
$T$ the temperature of the emitted neutrino flavor.
Each flavor is characterized by its
own temperature as follows:
$$T=8 \mbox { MeV for } \nu_{\mu},\nu_{\tau},\tilde{\nu}_{\mu}, \tilde{\nu}_{\tau}
\mbox{ and } T=5 ~(3.5)\mbox{ MeV for } \tilde{\nu}_e ~(\nu_e)$$
The constant $\Lambda$ is determined by the requirement that the distribution yields the total energy of each
neutrino species.
$$U_{\nu}=\frac{\Lambda T}{J}\int_0^{\infty } dx \frac{x^5}{1+e^x}\Rightarrow \Lambda=\frac{U_{\nu}}{T}$$
We will further assume that  $U_{\nu}=0.5 \times 10^{53}$ ergs
per neutrino flavor. Thus one finds:
$$\Lambda=0.89\times 10^{58}~(\nu_e),~~0.63\times 10^{58}~(\tilde{\nu}_e)~,0.39\times 10^{58}
\mbox{ (all other flavors)}$$
The emitted neutrino spectrum is shown in Fig. \ref{supernovasp}.

\begin{figure}[!ht]
 \begin{center}
% \rotatebox{90}{\hspace{1.0cm} {$F_{coh}$}}
\includegraphics[scale=0.5]{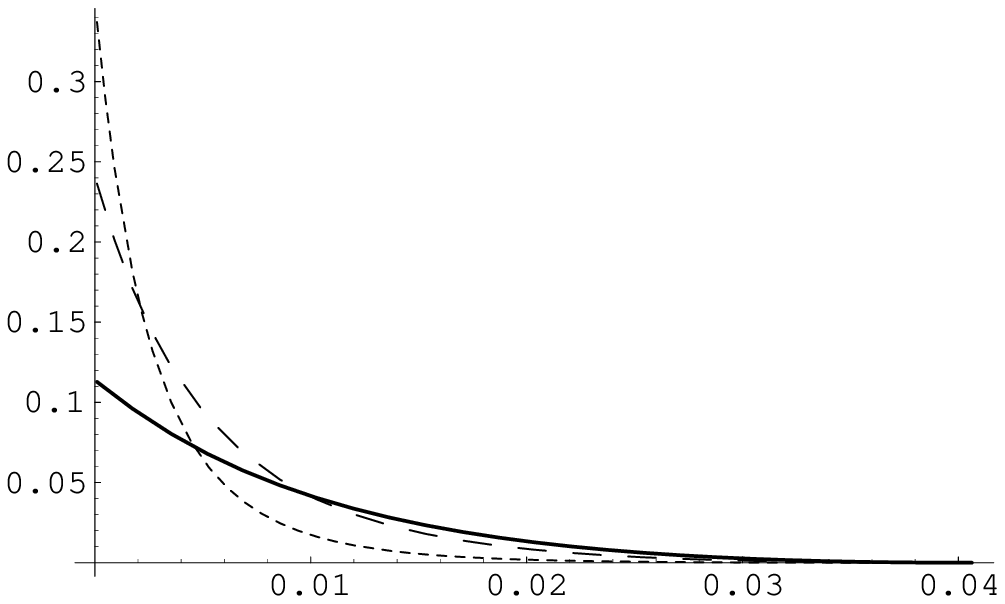}
\hspace*{0.0cm}\tiny{$T_A \rightarrow$ MeV}
\includegraphics[scale=0.5]{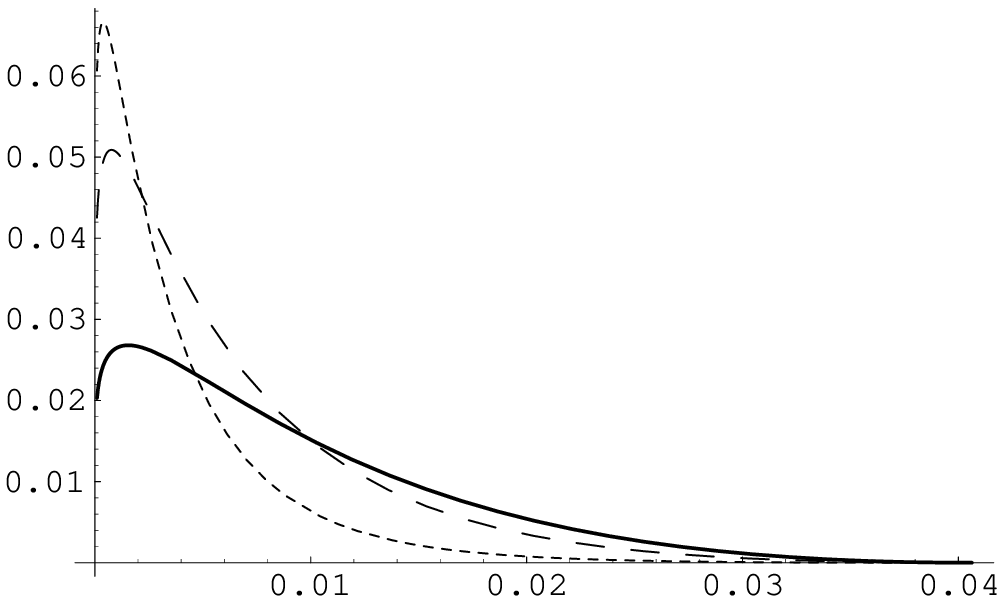}
\hspace*{0.0cm}\tiny{$T_A \rightarrow$ MeV}
 \caption{The differential event rate as a function of the recoil energy $T_A$, in arbitrary units, for
Xe. On  the left we show the results without quenching, while on the right the quenching factor is
included. We notice that the effect of quenching is more prevalent at low energies. The notation
for each neutrino species is
the same as in Fig. \ref{supernovasp}}
 \label{fig:difr131}
 \end{center}
  \end{figure}

  The differential event rate (with respect to the recoil energy) is proportional to the quantity:
\beq
\frac{dR}{dT_A}=\frac{\lambda (T)}{J}\int_0^{\infty } dx
F_{coh}(A,T_A,xT) \frac{x^4}{1+e^x}
\label{dRdT}
\eeq
with $\lambda(T)=(0.89,0.63,0.39)$  for $\nu_e,\tilde{\nu}_e$ and all other flavors respectively.
This is shown  in Figs. \ref{fig:difr131} and \ref{fig:difr131}.
%   \begin{figure}[!ht]
% \begin{center}
% \rotatebox{90}{\hspace{1.0cm} {$F_{coh}$}}
%\includegraphics[scale=0.5]{plotdifp40.eps}
%\hspace*{0.0cm}\tiny{$T_A \rightarrow$ MeV}
%\includegraphics[scale=0.5]{quplotdifp40.eps}
%\hspace*{0.0cm}\tiny{$T_A \rightarrow$ MeV}
% \caption{The same as in Fig. \ref{fig:difr131} for the Ar target.}
% \label{fig:difr40}
% \end{center}
%  \end{figure}
The total number of expected events for each neutrino species can be cast in the form:
\begin{eqnarray}
\mbox{No of events}&=&\tilde{C}_{\nu} (T) h(A,T,(T_A)_{th}),
\nonumber\\
\\h(A,T,(T_A)_{th})
&=&\frac{F_{fold}(A,T,(T_A)_{th})}{F_{fold}(40,T,(T_A)_{th})}
\label{events}
\end{eqnarray}
with
\begin{eqnarray}
F_{fold}(A,T,(T_A)_{th})&=&\frac{A}{J}\int_{(T_A)_{th}}^{(T_A)_{max}}\frac{dT_A}{1MeV}\times
\nonumber\\
& &\int_0^{\infty } dx F_{coh}(A,T_A,xT) \frac{x^4}{1+e^x}
\label{events1}
\end{eqnarray}
and
\beq
\tilde{C}_{\nu}(T)=\frac{G^2_F m_N1MeV}{2 \pi} \frac{N^2}{4}\Lambda (T)\frac{1}{4 \pi D^2}\frac{PV}{kT_0}
\label{C1}
\eeq
Where $k$ is Boltzmann's constant, $P$ the pressure, $V$ the volume, and $T_0$ the temperature of the gas.

Summing over all the neutrino species we can write:
\beq
\mbox{No of events}=C_{\nu} \frac{K(A,(T_A)_{th})}{K(40,(T_A)_{th})}Qu(A)
\label{sumevents}
\eeq
with
\beq
C_{\nu}=153  \left ( \frac{N}{22} \right )^2 \frac{U_{\nu}}{0.5\times 10^{53}ergs}
\left ( \frac{10kpc}{D}\right )^2 \frac{P}{10Atm}
\left[ \frac{R}{4m}\right]^3 \frac{300}{T_0}
\label{C2}
\eeq
%with
%\beq
%C=3.7 \times 10^{3}~(\nu_e),2.6\times 10^{3}~(\tilde{\nu}_e)~,1.6\times 10^{3}\mbox{ (other flavors)}
%\label{C3}
%\eeq
%The function $F_{fold}(A)$ scales as $A^{-1}$ so it is more convenient to write:
%\beq
%\mbox{No of events}=C_{\nu}r(A)~,~r(A)=\frac{h(40)}{h(A)}
%\label{events2}
%\eeq
%with
%\beq
%h(A)=\left[ F_{fold}(A) \right ]^{-1}
%\label{events3}
%\eeq
%The function $h(A,(T_A)_{th})$ can be written as follows:
%\beq
%h(A,(T_A)_{th})=r(A)K(A,(T_A)_{th})
%\label{factor}
%\eeq
%In the above expression $r(A)$ is a kinematical parameter depending on the nuclear mass number,
%which is essentially unity.

\begin{figure}[!ht]
 \begin{center}
 \rotatebox{90}{\hspace{-0.0cm} {\tiny $K(A,(T_A)_{th})\rightarrow $}}
 %\hspace{0.0cm}${\tiny K(T_A)_{th}} \rightarrow$MeV
\includegraphics[scale=0.45]{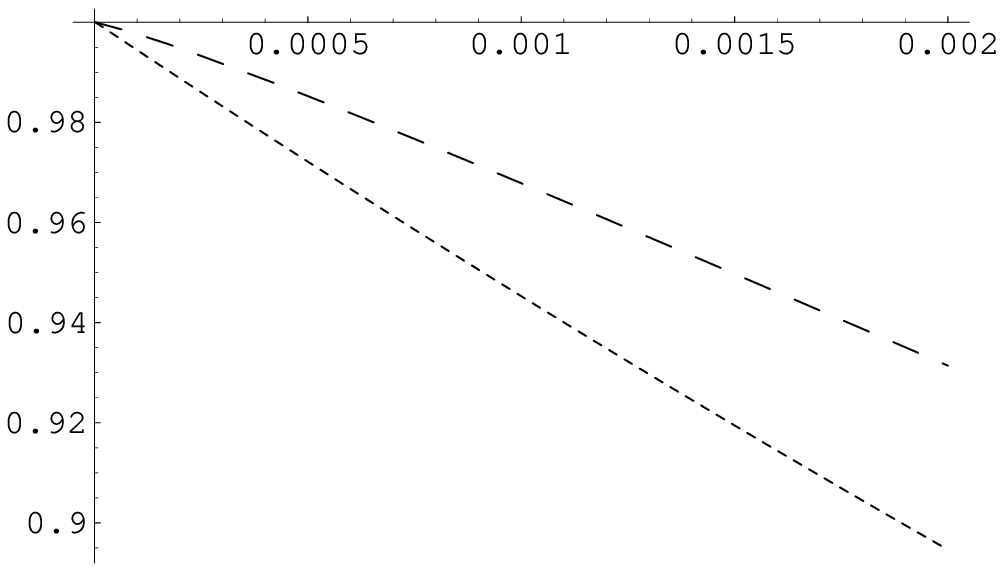}
 \hspace*{0.0cm}\tiny{$(T_A)_{th} \rightarrow$MeV }
  \rotatebox{90}{\hspace{-0.0cm} {\tiny $K(A,(T_A)_{th})\rightarrow $}}
\includegraphics[scale=0.45]{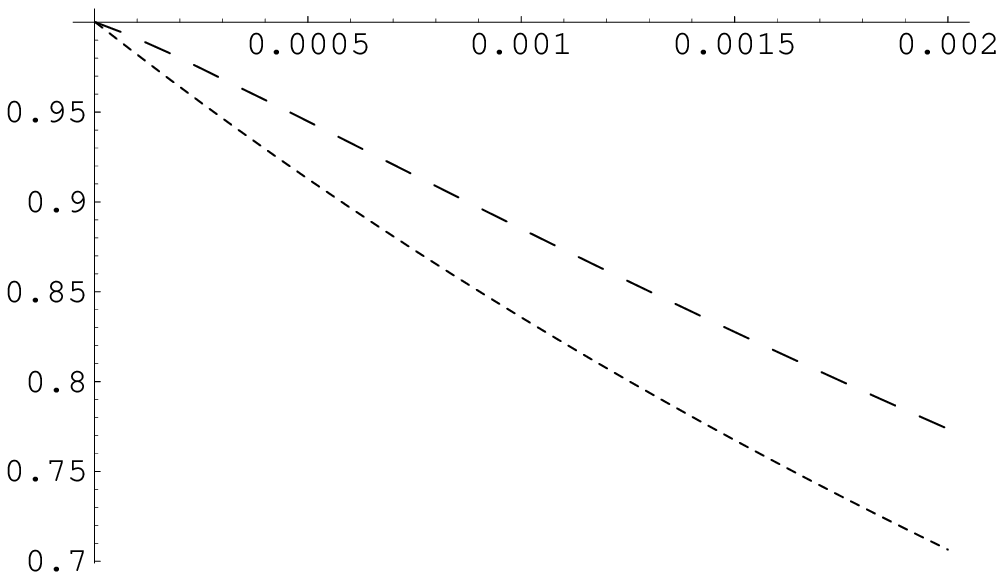}
\hspace*{0.0cm}\tiny{$(T_A)_{th} \rightarrow$MeV }
 \caption{The function $K(A,(T_A)_{th})$ versus $(T_A)_{th}$ for the target Ar on the left and
Xe on the right. The short and long dash correspond to no quenching and quenching factor respectively.
One sees that the effect of quenching is less pronounced at higher thresholds.
The differences appear small, since we present here only  the ratio of the rates to that at zero
threshold. The effect of quenching at some specific threshold energy is not shown here.
For a threshold energy of $100$ eV the rates are quenched by factors of $3$ and $3.5$ for Ar and Xe
 respectively (see Eq. (\ref{quench2}).}
 \label{fig:Kqu}
 \end{center}
  \end{figure}

% $(T_A)_{th}=0$, while
 $K(A,(T_A)_{th})$
 is  the rate at a given threshold energy divided by that at zero threshold. It depends
 on the threshold
energy, the assumed quenching factor and the nuclear mass number. It is unity at $(T_A)_{th})=0$.
%The function $r(A)$ is plotted in \ref{fig:ratio}. It is seen that  it can be well approximated by unity.\\
 From the above equation we find
that, ignoring quenching, the following expected number of events:
\beq
1.25,~31.6,~153,~614,~1880\mbox{ for He, Ne, Ar, Kr and Xe}
\label{allrates}
\eeq
respectively. For other possible targets the rates can be found by the above formulas or interpolation.\\
 The quantity $Qu(A)$ is a factor less than one multiplying the total rate, assuming a  threshold energy
  $(T_A)_{th}=100$eV, due to the quenching. The idea of quenching is introduced, since, for low emery recoils,
 only a fraction of the total deposited energy goes into
 ionization. The ratio of the amount of ionization induced in the gas due to nuclear recoil to the amount of ionization
 induced
by an electron of the same kinetic energy is referred to as a quenching factor $Q_{fac}$. This factor depends mainly on the
detector material, the recoiling energy as well as the process considered \cite{SIMON03}.
 In our estimate of $Qu(T_A)$ we assumed a quenching factor of the following empirical form motivated by the Lidhard
theory \cite{SIMON03}-\cite{LIDHART}:
\beq
Q_{fac}(T_A)=r_1\left[ \frac{T_A}{1keV}\right]^{r_2},~~r_1\simeq 0.256~~,~~r_2\simeq 0.153
\label{quench1}
\eeq
Then the parameter $Qu(A)$ takes the values:
\beq
0.49,~0.38,~0.35,~0.31,~0.29\mbox{ for He, Ne, Ar, Kr and Xe}
\label{quench2}
\eeq
respectively. The effect of quenching is larger in the case of  heavy targets, since, for a given neutrino energy, the energy of
the recoiling nucleus is smaller. Thus the number of expected events for Xe assuming a threshold energy of
$100$ eV is reduced to about 560.\\
  The effect of quenching is exhibited in Fig \ref{fig:Kqu}  for the two interesting targets
Ar and Xe.
 We should mention that it is of paramount importance to experimentally measure the quenching factor. The
 above estimates were based on the assumption of a pure gas.
 Such an effect will lead to an increase  in the quenching factor and needs be measured.

\section{Conclusions}
 In the present study it has been shown that it is quite simple to detect typical supernova
neutrinos in our galaxy, provided that such a supernova explosion takes place (one explosion every 30 years is estimated \cite{SOLBERG}). The idea is to employ a small size spherical TPC detector filled with a high
pressure noble gas. An enhancement of the neutral current component is achieved via the coherent
effect of all neutrons in the target. Thus employing, e.g., Xe at $10$ Atm, with a feasible threshold energy
of about $100$ eV in the detection the recoiling nuclei,
 one expects between $600$ and $1900$ events, depending on the quenching factor.
We believe that networks of such dedicated detectors, made out of simple, robust and cheap technology,
 can be simply managed by an international scientific consortium and operated by students. This network
 comprises a system, which can be maintained
for several decades (or even centuries). This is   is a key point towards being able to observe
 few galactic supernova explosions.

acknowledgments:  One of the authors (JDV) is indebted for support and hospitality to the OMEG05 organizing
committee during the OMEG05 conference and to Professor Hiroshi Toki during a visit to RCNP.
%.......................................................................

\end{document}